\documentclass[twocolumn,floatfix,aps,nofootinbib]{revtex4}

\usepackage{amsmath}
\usepackage{amssymb}
\usepackage{amscd}
\usepackage{mathrsfs}
\usepackage{graphicx}
\usepackage{subfigure}
\usepackage{slashed}
\usepackage{mathbbol}
\usepackage{subfigure}

\begin{document}

\title{ATLAS and CMS hints for a mirror Higgs boson}
\author{Robert Foot}\email{rfoot@unimelb.edu.au}
\affiliation{ARC Centre of Excellence for Particle Physics at the Terascale, School of Physics, The University of Melbourne, Victoria 3010, Australia}
\author{Archil Kobakhidze}\email{archilk@unimelb.edu.au}
\affiliation{ARC Centre of Excellence for Particle Physics at the Terascale, School of Physics, The University of Melbourne, Victoria 3010, Australia}
\author{Raymond R. Volkas}\email{raymondv@unimelb.edu.au}
\affiliation{ARC Centre of Excellence for Particle Physics at the Terascale, School of Physics, The University of Melbourne, Victoria 3010, Australia}

\begin{abstract}
ATLAS and CMS have provided hints for the existence of a Higgs-like particle with mass of about 144 GeV
with production cross section into standard decay channels which is about $50\%$ that of the standard model Higgs
boson.  We show that this $50\%$ suppression is exactly what the mirror matter model predicts when the two
scalar mass eigenstates, each required to be maximal admixtures of a standard and mirror-Higgs boson,
are separated in mass by more than their decay widths but less than the experimental resolution.
We discuss prospects for the future confirmation of this interesting hint for non-standard Higgs physics.

\end{abstract}

\maketitle

The ATLAS \cite{atlas} and CMS \cite{cms} experiments 
at the Large Hadron Collider (LHC) have achieved major milestones this year, already excluding a large range
of parameter space for the standard model Higgs boson.
These data, in combination with LEP and precision electroweak data \cite{pdg}, restricted the Higgs boson mass to the range $115 - 146$ GeV. Within this allowed mass range, there is an interesting hint in the ATLAS and CMS data for a Higgs-like particle
with a mass of about $144$ GeV at about the $2 \sigma$ confidence level for each experiment. However as noted in 
Ref.~\cite{recent}
the size of the signal is roughly a factor of two less than that expected for the standard model
Higgs boson: $\sigma/\sigma_{\rm SM Higgs} \approx 0.5 \pm 0.2$,
which has lead the authors of Ref. \cite{recent} to speculate that these experiments
might be seeing a Higgs-like particle with a $\sim 50\%$ branching fraction into invisible
decay modes.  The discovery of a Higgs-like particle with invisible decay modes would be revolutionary, because
it would be tantamount to the discovery of a hidden sector, quite probably related to dark matter, in addition to
the origin of electroweak symmetry breaking.

A number of different theoretical proposals for invisibly decaying Higgs bosons have been
studied in the literature; see, for example,  \cite{inv}. One of the earliest such proposals was 
discussed in the context of the mirror matter model \cite{flv,flv2,sasha,hall,chin}. In fact,
it was pointed out some time ago \cite{flv2} that each of the two physical Higgs particles 
in that model will
decay invisibly $50\%$ of the time provided that there is significant Higgs--mirror-Higgs mass mixing,
and the production cross-section for each of them will be half that of a standard Higgs boson.
Thus each state has an effective production cross-section to visible particles of $25\%$.
The recent ATLAS and CMS results motivate a reanalysis of the implications of Higgs--mirror-Higgs mass mixing.
Indeed, we shall point out there is a region of parameter space that was missed in the earlier 
studies \cite{flv,flv2,sasha,hall,chin}, but which provides a possible explanation of the new data.  In this
previously unidentified regime, the two Higgs particles add incoherently thus reproducing the desired $50\%$
figure for the production of standard particles.

The mirror matter model \cite{flv} postulates the existence of a hidden sector described by a Lagrangian
exactly isomorphic to the standard model Lagrangian, except with the role of left-
and right-handed fermion fields interchanged in the hidden sector.
The Lagrangian is thus,
\begin{eqnarray}
\mathcal{L} = \mathcal{L}_{SM} (e, \nu, u, d, ...) + \mathcal{L}_{SM} (e', \nu', u', d',...)
+ \mathcal{L}_{\rm mix}.
\label{lag}
\end{eqnarray}
The
gauge symmetry of the theory is $G_{SM}\otimes G_{SM}$ where
$G_{SM} = SU(3)_c \otimes SU(2)_L \otimes U(1)_Y$. In addition, exact improper space-time
symmetries can be defined which provides an interesting theoretical motivation
for the theory. In fact, the Lagrangian (\ref{lag}) contains an unbroken parity symmetry which maps each ordinary 
particle to its mirror partner. The mirror particles, denoted by primes in Eq.~(\ref{lag}),
also provide stable mirror atoms and ions which are viable candidates for
non-baryonic dark matter \cite{dm1,dm}, in a way consistent with the positive results from the DAMA \cite{dama} and 
CoGeNT \cite{cogent} 
direct detection experiments \cite{fr}.
The Lagrangian $\mathcal{L}_{\rm mix}$ describes renormalisable mixing terms which are consistent with 
the symmetries of the theory,
\begin{eqnarray}
\mathcal{L}_{\rm mix} = {\epsilon \over 2} F^{\mu \nu} F'_{\mu \nu} + 
\lambda_2 \phi^{\dagger} \phi \phi'^{\dagger} \phi',
\label{1}
\end{eqnarray}
where $F^{\mu \nu}$ ($F'^{\mu \nu}$) is the $U(1)_Y$ [$U(1)'_Y$] field strength tensor and
$\phi$ ($\phi'$) is the ordinary (mirror) Higgs doublet.
The $\lambda_2 \phi^{\dagger} \phi \phi'^{\dagger} \phi'$ term 
provides the Higgs--mirror-Higgs mass mixing that can lead to non-standard Higgs boson signatures at the LHC.

The complete Higgs potential is
\begin{eqnarray}
V(\phi, \phi') & = & -\mu^2 \left( \phi^{\dagger} \phi + \phi'^{\dagger}\phi' \right) + 
\lambda_1 \left[(\phi^{\dagger} \phi)^2 + (\phi'^{\dagger} \phi')^2\right] \nonumber\\
& + & \lambda_2 \phi^{\dagger}\phi \phi'^{\dagger} \phi'.
\end{eqnarray}
If $\lambda_1 > 0$ and $|\lambda_2| < 2\lambda_1$, then the parity conserving minimum
$\langle \phi \rangle = \langle \phi' \rangle = 
(0, {v \over \sqrt{2}})^T$ occurs, with $v = \sqrt{2\mu^2 \over 2\lambda_1 + \lambda_2} \simeq 246$ GeV.
There are two mass eigenstate Higgs fields, $H_1$ and $H_2$, which are maximal combinations
of the weak eigenstates:
\begin{eqnarray}
H_1 = {\phi_0 + \phi'_0 \over \sqrt{2}}, \ 
H_2 = {\phi_0 - \phi'_0 \over \sqrt{2}},
\end{eqnarray}
where $\phi_0$ and $\phi'_0$ are the real parts of the neutral components of $\phi$ and $\phi'$, respectively.
The maximal admixture is simply due to the fact that the mirror-parity eigenstates must also be mass eigenstates,
with $H_1 (H_2)$ having positive (negative) parity.
The Lagrangian, Eq.~(\ref{1}), when rewritten in terms of these mass eigenstates,
reveals that each of them couples to the standard particles with coupling constants that are
$1/\sqrt{2}$ times those of the corresponding standard model Higgs boson.  This means
that the production cross-section for each mass eigenstate is $50\%$ that for the standard Higgs boson.

The states $H_1$ and $H_2$ similarly couple to mirror particles, implying   
that they decay invisibly $50\%$ of the time provided that their mass difference is large enough.
The masses of $H_{1,2}$ are given by
\begin{eqnarray}
m_{H_1}^2 = 2v^2 \left(\lambda_1 + {\lambda_2 \over 2}\right), \ 
m_{H_2}^2 = 2v^2\left(\lambda_1 - {\lambda_2 \over 2}\right).
\end{eqnarray}
The important mass difference parameter, $|m_{H_1} - m_{H_2}|$,
is given by
\begin{eqnarray}
|m_{H_1} - m_{H_2}| &=& \sqrt{2} v \left| \sqrt{\lambda_1 + {\lambda_2 \over 2}} -
\sqrt{\lambda_1 - {\lambda_2 \over 2}}\right|,  \nonumber \\
&\simeq & {|\lambda_2| v \over \sqrt{2 \lambda_1}} \ {\rm for} \ |\lambda_2| \ll \lambda_1 \ .
\end{eqnarray}
This last regime, where the mass difference is not too large, 
will be of relevance for potentially explaining the ATLAS and CMS results.
Since the limit $\lambda_2 \to 0$ corresponds to the decoupling of the ordinary and mirror
sectors, this limit is technically natural and we therefore do not consider small values of $\lambda_2$
to be 'fine-tuned'.

The experimental implications depend on the value of $|m_{H_1} - m_{H_2}|$.
There are three regimes of interest. First, one can have $|m_{H_1} - m_{H_2}|$ greater than
the experimental Higgs mass resolution; 
in this case each mass eigenstate can be resolved. The cross-sections to visible channels depend
on whether or not the mass difference is large enough to kinematically allow $H_1 \to H_2 H_2$
(see Ref.~\cite{chin} for a study of this possibility).  For the situation where the mass difference is not
sufficiently large to allow this, the cross-sections to visible channels 
for $H_1$ and $H_2$ are each $25\%$ that of the standard model Higgs boson.
In the most sensitive channel for a 140 GeV Higgs boson, $H \to WW^* \to 2\ell 2\nu$, 
the experimental Higgs boson mass
resolution is very large, of order 30 GeV, because of the 
undetected neutrinos.  
For the golden channel, $H \to ZZ^* \to 4 \ell$, also important for a $\sim 140$ GeV
Higgs, the experimental resolution is much less, of order $5$ GeV.

This large experimental resolution width leads to the existence of the second regime.
This possibility, which was missed in the earlier studies \cite{flv,flv2,sasha,hall,chin},
is that $|m_{H_1} - m_{H_2}|$ is less than the experimental resolution, but
larger than the $H_{1,2}$ decay widths, $\Gamma_H$. For $m_{H} \sim 140$ GeV, 
$\Gamma_H \approx 8$ MeV \cite{width}.
In this case the two states cannot be resolved,
and the combined signal to visible channels is expected to be about $50\%$ that of the standard model Higgs
\footnote{Strictly, a $50\%$ signal reduction is only expected provided that $|m_{H_1} - m_{H_2}|$ is sufficiently
small so that the cross-section of the process involving $H_1$ is approximately the same as that
involving $H_2$. For $|m_{H_1} - m_{H_2}| \stackrel{<}{\sim} 5 $ GeV this should certainly be the case.
For larger values of $|m_{H_1} - m_{H_2}|$ significant deviations from $50\%$ are possible, depending on the process.}.
This parameter range can thus nicely explain the factor of two reduction for the cross section 
of the tentative $\sim 140$ GeV
Higgs signal. Obviously, further data will be needed to confirm such a signal.

The third possibility occurs when $|m_{H_1} - m_{H_2}| \lesssim \Gamma_{H}$.
In this case the weak eigenstate, $\phi_0$, is produced and begins to maximally oscillate
into the mirror state $\phi'_0$.
The oscillation probability is
\begin{eqnarray}
P(\phi_0 \to \phi'_0) = \sin^2 {t \over 2t_{\rm osc}}
\end{eqnarray}
where $t_{\rm osc} = 1/|m_{H_1} - m_{H_2}|$ in the non-relativistic 
limit.\footnote{We use units where $\hbar = c = 1$.}
The invisible branching fraction of the Higgs boson in this case corresponds to the   
average oscillation probability to the mirror state, which is
\begin{eqnarray}
\langle P(\phi_0 \to \phi'_0) \rangle &=& \Gamma_H \int_0^{\infty} e^{-\Gamma_H t} \sin^2 {t \over 2t_{\rm osc}}
dt \nonumber \\
&=& {1 \over 2} \left( {1 \over 1 + \Gamma_H^2 t_{\rm osc}^2}\right).
\end{eqnarray}
Evidently the branching fraction to invisible channels is always less than $50\%$ in this last regime.
Note that the cross-section into visible channels is reduced by the factor $f$, where
\begin{eqnarray}
f &=& 1 - \langle P(\phi_0 \to \phi'_0)\rangle \nonumber \\
&=& {1 \over 2} + {1 \over 2} \left( {\Gamma_H^2 t_{\rm osc}^2 \over 1 + \Gamma_H^2 t_{\rm osc}^2} \right).
\end{eqnarray}
Observe that in the limit where $|m_{H_1} - m_{H_2}| \to 0$ (or 
equivalently $\lambda_2 \to 0$), $f \to 1$, and the signal becomes indistinguishable from that of the 
standard Higgs boson case.

If the LHC is indeed seeing a Higgs-like particle with mass $\sim 140$ GeV 
and the cross-section is half of the Standard Model Higgs cross section, 
then roughly $5-10$ fb$^{-1}$ of data would be needed to confirm this at $\sim 5\sigma$ level. 
The invisible decay channels of $H_{1,2}$ can also in principle be detected at the LHC via processes such as
$pp \to Z^* \to Z H_{1,2}$ leading to $Z$ plus missing energy signature. 
However, the low cross section of this process and significant backgrounds mean that it might take
some time to observe this signature (see, for example, the discussions in Ref.~\cite{inv}).
In fact, the study \cite{atlas2} suggests that around $30$ fb$^{-1}$
of integrated luminosity will be needed to detect such a Higgs boson via missing energy signatures.

The compatibility of significant Higgs--mirror-Higgs mixing with early universe cosmology 
was studied in Ref.~\cite{sasha}, where it was shown that all cosmological bounds could
be evaded within inflationary scenarios if the reheating temperature was sufficiently low.
For a Higgs mass around 140 GeV the maximum allowed reheating temperature after inflation, $T_{\rm max}$,
was estimated to be around $T_{\rm max} \sim 1 - 100$ GeV, depending on $\lambda_2$.  If such a low
reheating temperature were to be established, it would rule out all high-temperature mechanisms for
baryogenesis.

Indications of a mirror sector from LHC Higgs physics will obviously further strengthen the case
that dark matter in the Universe is composed of mirror particles. 
Mirror matter can play the role of dark matter in the Universe provided
a number of conditions are met. These include:
a) The successful Big Bang  nucleosynthesis and  large scale structure formation suggests that   
$T' << T$ and $n_{b'} \approx 5n_b$,  where  $n_{b'}$ [$n_b$] is the  number density of mirror [ordinary] baryons and $T'$  [$T$] is the temperature of the mirror [ordinary] particles in the early Universe. Such initial conditions cam be generated in models of the early Universe \cite{dm1}. b) Observations of the Bullet cluster \cite{Clowe:2006eq} suggest that mirror dark matter needs to be 
predominantly confined into galaxies, within clusters, to be consistent with observations.
This is because mirror dark matter is self-interacting.
c) Mapping of the dark matter density outside galactic halos in clusters
of galaxies using gravitational lensing suggest that dark matter
is distributed smoothly on scales greater than of order 10 kpc \cite{Tyson:1998vp}. This
suggests numerous (probably small) dark galaxies within clusters.
d) Mirror dark matter is dissipative, which suggests the existence of a  
substantial galactic heat source to replace the energy lost due to radiative cooling.
The most plausible candidate is the energy from ordinary core collapse supernovae 
which can transfer the required energy into the mirror sector if kinetic
mixing interaction exists with $\epsilon \sim 10^{-9}$ \cite{sn}.
This heat source is anisotropic in spiral galaxies (given that the supernova originate predominately
within the galactic disk and bulge) which might explain the deviations from
perfect spherical halos which are necessary to agree with observations.

To summarise, we have re-examined the implications of Higgs-mirror-Higgs mass mixing for the
experiments at the LHC.
There are three interesting parameter regimes, which depend on the mass difference $|m_{H_1} - m_{H_2}|$.
Current ATLAS and CMS data provide a tantalising hint for a Higgs-like particle with mass $\sim 144$ GeV
and production cross section to visible channels around $50\%$ that
of the standard model Higgs boson. We have shown here that this hint, if confirmed, could be 
simply explained within the mirror matter model if $|m_{H_1} - m_{H_2}|$ is less than the experimental
resolution ($\sim 5-30$ GeV, depending on the process) but greater than the Higgs width ($\sim 8$ MeV for $m_H \sim 140$ GeV). 
A $5\sigma$ discovery is possible with $5-10$ fb$^{-1}$ of data.  Confirmation of the result would have
startling implications for physics beyond the standard model, through the discovery of a hidden sector
isomorphic to the standard model and a strong constraint on the thermal history of the universe.

\vskip 1cm

\noindent
{\bf Acknowledgments}

\vskip 0.3cm

\noindent
We thank Geoff Taylor for comments on the manuscript. This work was supported in part by the Australian Research Council.


\begin{thebibliography}{999}

\bibitem{atlas}
The ATLAS Collaboration, ``Update of the Combination of Higgs Boson Searches in 1.0 to 2.3 1/fb of pp 
Collisions Data Taken at $\sqrt{s}$ = 7
TeV with the ATLAS Experiment at the LHC,"  ATLAS-CONF-2011-135,
https://atlas.web.cern.ch/Atlas/GROUPS/PHYSICS/
CONFNOTES/ATLAS-CONF-2011-135/ATLAS-CONF-2011-135.pdf



\bibitem{cms}
The CMS Collaboration, ``Search for standard model Higgs boson in pp collisions at $\sqrt{s}$ = 7 TeV and integrated luminosity up to 1.7
1/fb,"
CMS-PAS-HIG-11-022, 
http://cdsweb.cern.ch/record/1376643/files/HIG-11-022-pas.pdf


\bibitem{pdg}
K. Nakamura {\it et al}. (particle Data Group), J. Phys. G37, 075021 (2010).


\bibitem{recent}
M. Raidal and A. Strumia, arXiv: 1108.4903.


\bibitem{inv}
O.~J.~P.~Eboli and D.~Zeppenfeld, Phys.\ Lett.\  B {495}, 147 (2000)
[arXiv:hep-ph/0009158];
R.~M.~Godbole, M.~Guchait, K.~Mazumdar, S.~Moretti and D.~P.~Roy,
Phys.\ Lett.\  B {571}, 184 (2003) [arXiv:hep-ph/0304137];
M.~Duhrssen, S.~Heinemeyer, H.~Logan, D.~Rainwater, G.~Weiglein and D.~Zeppenfeld,
Phys.\ Rev.\  D {70}, 113009 (2004) [arXiv:hep-ph/0406323];
H.~Davoudiasl, T.~Han and H.~E.~Logan, Phys.\ Rev.\  D {71}, 115007 (2005) [arXiv:hep-ph/0412269];
V. Barger {\it et al}.,
Phys.\ Rev.\ D {77}, 035005 (2008) [arXiv 0706.4311];
C. Englert, T. Plehn, D. Zerwas and P. M. Zerwas, arXiv: 1106.3097; Y.~Mambrini, [arXiv:1108.0671 [hep-ph]].

\bibitem{flv}
R. Foot, H. Lew and R. R. Volkas,
Phys.\ Lett.\ B272, 67 (1991).

\bibitem{flv2}
R. Foot, H. Lew and R. R. Volkas,
Mod.\ Phys.\ Lett.\ A7, 2567 (1992).


\bibitem{sasha}
A. Yu. Ignatiev and R. R. Volkas, Phys.\ Lett.\ B487, 294 (2000) [arXiv: hep-ph/0005238].

\bibitem{hall}
R. Barbieri, T. Gregoire, L. J. Hall, hep-ph/0509242.

\bibitem{chin}
W. Li, P. Yin and S. Zhu, Phys.\ Rev\. D76, 095012 (2007) [arXiv:0709.1586].

\bibitem{dm1}
H. M. Hodges, Phys.\ Rev.\ D47, 456 (1993);
Z. Berezhiani, D. Comelli and F. L. Villante, Phys.\ Lett.\ B503, 362 (2001)
[hep-ph/0008105];
R. Foot and R. R. Volkas, Phys.\ Rev.\ D68, 021304 (2003) [hep-ph/0304261];
Phys.\ Rev.\ D69, 123510 (2004) [hep-ph/0402267]. 

\bibitem{dm}
L. Bento and Z. Berezhiani, Phys.\ Rev.\ Lett.\ 87, 231304 (2001) [hep-ph/0107281];
A. Yu. Ignatiev and R. R. Volkas, Phys.\ Rev\. D68, 023518 (2003) [hep-ph/0304260]; 
R. Foot and R. R. Volkas, Phys.\ Rev.\ D70, 123508 (2004)
[arXiv: astro-ph/0407522]; Z. Berezhiani, P. Ciarcelluti,
D. Comelli and F. L. Villante, Int.\ J. Mod.\ Phys.\ D14, 107 (2005) [astro-ph/0312605];
P. Ciarcelluti, Int.\ J. Mod.\ Phys.\ D14, 187 (2005) [astro-ph/0409630];
Int.\ J. Mod.\ Phys.\ D14, 223 (2005) [astro-ph/0409633]. For pioneering work
see S. I. Blinnikov and M. Yu. Khlopov, Sov.\ J. Nucl.\ Phys.\ 36, 472 (1981);
Sov.\ Astron.\ 27, 371 (1983).

\bibitem{dama}
R.~Bernabei {\it et al.}, Eur.\ Phys.\ J.\  C67, 39 (2010) [arXiv:1002.1028]; 
Eur.\ Phys.\ J.\  C56, 333 (2008) [arXiv:0804.2741];  Riv.\ Nuovo Cim.\  26N1, 1 (2003) [astro-ph/0307403].
  
\bibitem{cogent}
C.~E.~Aalseth {\it et al.}, arXiv:1106.0650; Phys.\ Rev.\ Lett.\  106, 131301 (2011) [arXiv:1002.4703].

\bibitem{fr}
R. Foot, Phys. Lett. B703, 7 (2011) [arXiv: 1106.2688];
Phys.\ Rev.\ D82, 095001 (2010) [arXiv: 1008.0685] and references therein.

\bibitem{width}
See e.g.
M. Spira and P. M. Zerwas, Lect.\ Notes.\ Phys.\ 512, 161 (1998)
[arXiv: hep-ph/9803257].

\bibitem{atlas2}
ATLAS note ATL-PHYS-PUB-2009-061,
http://cdsweb.
cern.ch/record/1174275/files/ATL-PHYS-PUB-2009-061.pdf.

\bibitem{Clowe:2006eq}
  D.~Clowe, M.~Bradac, A.~H.~Gonzalez, M.~Markevitch, S.~W.~Randall, C.~Jones and D.~Zaritsky,
  Astrophys.\ J.\  {\bf 648}, L109 (2006)
  [arXiv:astro-ph/0608407].

\bibitem{Tyson:1998vp} See,e.g., 
  J.~A.~Tyson, G.~P.~Kochanski and I.~P.~Dell'Antonio,
  Astrophys.\ J.\  {\bf 498}, L107 (1998)
  [arXiv:astro-ph/9801193].

\bibitem{sn}
R. Foot and R. R. Volkas, Phys.\ Rev.\ D70, 123508 (2004)
[arXiv: astro-ph/0407522]. See also, 
  R.~Foot and Z.~K.~Silagadze,
  Int.\ J.\ Mod.\ Phys.\  D {\bf 14}, 143 (2005)
  [arXiv:astro-ph/0404515].


\end{thebibliography}
\end{document}